# A Data-Driven Analysis of Workers' Earnings on Amazon Mechanical Turk

Kotaro Hara[1,2], Abi Adams[3], Kristy Milland[4,5],
Saiph Savage[6], Chris Callison-Burch[7], Jeffrey P. Bigham[1]
[1]Carnegie Mellon University, [2]Singapore Management University, [3]University of Oxford
[4]McMaster University, [5]TurkerNation.com, [6]West Virginia University, [7]University of Pennsylvania
kotarohara@smu.edu.sg

## ABSTRACT
A growing number of people are working as part of on-line crowd work. Crowd work is often thought to be low wage work. However, we know little about the wage distribution in practice and what causes low/high earnings in this setting. We recorded 2,676 workers performing 3.8 million tasks on Amazon Mechanical Turk. Our task-level analysis revealed that workers earned a median hourly wage of only ~$2/h, and only 4% earned more than $7.25/h. While the average requester pays more than $11/h, lower-paying requesters post much more work. Our wage calculations are influenced by how unpaid work is accounted for, *e.g.*, time spent searching for tasks, working on tasks that are rejected, and working on tasks that are ultimately not submitted. We further explore the characteristics of tasks and working patterns that yield higher hourly wages. Our analysis informs platform design and worker tools to create a more positive future for crowd work.

## ACM Classification Keywords
H.5.m. Information interfaces and presentation (e.g., HCI): Miscellaneous.

## INTRODUCTION
Crowd work is growing [31,46]. A report by Harris and Krueger states that 600k workers participate in the online gig economy and the number is growing rapidly [31]. Crowdsourcing does not just enable novel technologies (*e.g.,* human-powered word processing and assistive technologies [5,6]) that we create in the HCI community, but also facilitates new ways of working. Its remote and asynchronous work style, unbounded by time and location, is considered to extend the modern office work [44,46,47], enabling people with disabilities, at-home parents, and temporarily out-of-work engineers to work [1,4,39,46,65].

Yet, despite the potential for crowdsourcing platforms to extend the scope of the labor market, many are concerned that workers on crowdsourcing markets are treated unfairly [19,38,39,42,47,59]. Concerns about low earnings on crowd work platforms have been voiced repeatedly. Past research has found evidence that workers typically earn a fraction of the U.S. minimum wage [34,35,37–39,49] and many workers report not being paid for adequately completed tasks [38,51]. This is problematic as income generation is the primary motivation of workers [4,13,46,49].

Detailed research into crowd work earnings has been limited by an absence of adequate quantitative data. Prior research based on self-reported income data (*e.g.,* [4,34,49]) might be subject to systemic biases [22] and is often not sufficiently granular to facilitate a detailed investigation of earnings dispersion. Existing data-driven quantitative work in crowdsourcing research has taken the employers' perspective [49] (*e.g.,* finding good pricing methods [36,50,61], suggesting effective task design for requesters [24,40]), or it characterizes crowdsourcing market dynamics [21,37]. Data-driven research on how workers are treated on the markets is missing.

This paper complements and extends the existing understanding of crowd work earnings using a data-driven approach. Our research focuses on Amazon Mechanical Turk (AMT), one of the largest micro-crowdsourcing markets, that is widely used by industry [34,48] and the HCI community, as well as by other research areas such as NLP and computer vision [15,45]. At the core of our research is an unprecedented amount of worker log data collected by the Crowd Workers Chrome plugin [14] between Sept 2014 to Jan 2017. Our dataset includes the records of 3.8 million HITs that were submitted or returned by 2,676 unique workers. The data includes task duration and HIT reward, which allows us to evaluate hourly wage rates—the key measure that has been missing from the prior data-driven research [21,40]—at an unprecedented scale.

We provide the first task-level descriptive statistics on worker earnings. Our analysis reveals that the mean and median hourly wages of workers on AMT are $3.13/h and $1.77/h respectively. The hourly wage distribution has a long-tail; the majority of the workers earn low hourly wages, but there are 111 workers (4%) who earned more than $7.25/h, the U.S. federal minimum wage. These findings reify existing research based on worker self-reports that estimate the typical hourly wage to be $1-6/h [4,34,49] and strongly supports the view that crowd workers on this platform are underpaid [34,38]. However, it is not that individual requesters are necessarily paying so little, as we found requesters pay $11.58/h on average. Rather, there is a group of requesters who post a large amount of low-reward HITs and, in addition, unpaid time spent doing work-related activities leads to the low wages. We quantify three sources of unpaid work that impact the hourly wage: (i) searching for tasks, (ii) working on tasks that are rejected, and (iii) working on tasks that are not submitted. If one ignores this unpaid work, our estimates of the median and mean hourly wages rise to $3.18/h and $6.19/h respectively.

Our data also enable us to go beyond existing quantitative studies to examine how effective different work and task-selection strategies are at raising hourly wages. Workers could employ the potential strategies to maximize their hourly wage while working on AMT. In the final section, we discuss the implications of our findings for initiatives and design opportunities to improve the working environment on AMT and crowdsourcing platforms in general.

## BACKGROUND

Many are concerned that workers on crowdsourcing markets are treated unfairly [19,38,39,42,47,59]. Market design choices, it is argued, systematically favor requesters over workers in a number of dimensions. The use of asymmetric rating systems makes it difficult for workers to learn about unfair requesters [2,38,60], while platforms rarely offer protection against wage theft or provide mechanisms for workers to dispute task rejections and poor ratings [4,47]. Platforms' characteristics such as pay-per-work [2] and treating workers as contractors [64] (so requesters are not bound to paying minimum wage [64,66]) also contribute to earnings instability and stressful working conditions [4,11].

Past research has found evidence that workers typically earn a fraction of the U.S. minimum wage [34,35,37–39,49] and many workers report not being paid for adequately completed tasks [38,49,51]. This is problematic as income generation is the primary motivation of workers [4,13,46,49]. Further, low wage rates and the ethical concerns of workers should be of importance to requesters given the association between poor working conditions, low quality, and high turnover [12,27,44].

To date, detailed research into crowd work earnings has been limited by an absence of adequate quantitative data. For instance, Martin *et al.* analyzed publicly available conversations on Turker Nation—a popular forum for workers—in an attempt to answer questions such as "how much do Turkers make?" [49]. While such analyses have provided important insights into how much the workers believe they earn, we cannot be sure if their earnings estimates are unbiased and representative.

Existing quantitative work in crowdsourcing research has taken the employers' perspective [49] (*e.g.,* finding good pricing methods [36,50,61], suggesting effective task design for requesters [24,40]) or it focuses on characterizing the crowdsourcing market dynamics [21,37]. Although important, data-driven research on how workers are treated on the crowdsourcing markets is missing. This paper complements and extends our existing understanding of crowd work earnings using a data-driven approach. The unprecedented amount of AMT worker log data collected by the Crowd Workers Chrome plugin [14] allows us to evaluate hourly wage rates at scale.

## TERMINOLOGY

Before presenting our formal analysis, we define a set of key terms necessary for understanding the AMT crowdsourcing platform. AMT was launched in 2008 and is one of the largest micro-task sites in operation today. The 2010 report by Ipeirotis noted that the most prevalent types on AMT are transcription, data collection, image tagging, and classification [37]. Follow-up work by Difallah *et al.* reaffirms these findings, although tasks like audio transcription are becoming more prevalent [21].

Each standalone unit of work undertaken by a worker on AMT is referred to as a *task* or *HIT*. Tasks are listed on custom webpages nested within the AMT platform, although some tasks require workers to interact with web pages outside of the AMT platform.

Tasks are issued by *requesters*. Requesters often issue multiple HITs at once that can be completed by different workers in parallel. A group of tasks that can be performed concurrently by workers is called a *HIT group*.

Requesters can require workers to possess certain *qualifications* to perform their tasks. For example, a requester could only allow workers with "> 95% HIT approval rate" to work on their tasks.

Workers who meet the required qualifications can *accept* HITs. Once workers complete a task, they *submit* their work for requesters to evaluate and either *approve* or *reject* the HITs. If a submitted task is approved, workers get a financial *reward*. If, however, a worker accepts a HIT but does not complete the task, the task is said to be *returned*.

## DATASET

We describe the tool that we used to collect task-level data on worker behavior and present basic statistics of the data.

### Crowd Worker Plugin

The data was collected using the Crowd Workers Chrome plugin [14]. The plugin was used by workers in an opt-in basis. The plugin was designed to disclose the effective hourly wage rates of tasks for workers, following design suggestions in [55]. It tracks what tasks and when workers

accept and submit/return, as well as other metadata about the HITs. More specifically, our dataset includes:

- User attributes such as *worker ID*s, *registration date*, *blacklisted requesters*, *"favorite" requesters*, and *daily work time goal*.
- HIT Group information such as *HIT Group ID*s, *titles*, *descriptions*, *keywords*, *reward*, and *requester IDs*, and any qualification requirements.
- For each HIT group, we have information on *HIT IDs*, submission status (*i.e., submitted vs. returned*), timestamps for HIT accept, submit, and return.
- Web page domains that the workers visited (though the scope was limited to predefined domains including mturk.com, crowd-workers.com, and a selected few AMT-related sites (*e.g.,* turkernation.com).
- A partial record of HIT *approval* and *rejection* status for submitted HITs. The plugin periodically polled the worker's AMT dashboard and scraped this data. As an approve/reject status is updated by the workers at their convenience rather than at a specified interval after task completion, we only have approval records for 29.6% of the HIT records.

Some important attributes are not recorded in our dataset. For instance, the plugin does not record fine-grained interactions, such as keystrokes and mouse movements. Though potentially useful in, for example, detecting active work, we did not collect them because they could contain personally identifiable information. Further, while the plugin records data about browsing on a set of predefined web sites, it does not track browsing history on all domains. The plugin does not collect the HTML contents of the HIT UIs. Thus, we do not have the "true" answers for tasks that workers performed, so we cannot compute task accuracy.

**Data Description**

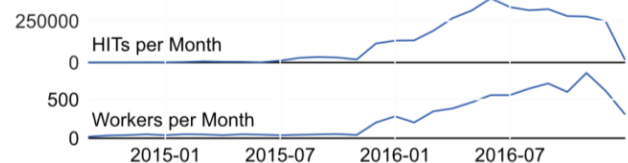

Figure 1. Line charts showing the transition in the number of active monthly users and HIT records.

The dataset consists of task logs collected from Sept 2014 to Jan 2017. There are 3,808,020 records of HITs from 104,939 HIT groups performed by 2,676 unique workers. The recorded HITs were posted by 20,286 unique requesters. Figure 1 shows the transition in the number of active monthly users and tracked HITs. We can see that the number of recorded HITs increased from December 2015 (N=114,129) and peaked on June 2016 (N=386,807). The data on January 2017 is small because the data was exported earlier in the month and the data for the full month was not collected. The number of unique monthly user started to increase from December 2015 (N=202), then peaked on November 2016 (N=842), indicating that the following analyses mainly reflect the activities from the end of 2015 to the end of 2016. To our knowledge, this is the largest AMT worker log data in existence that enables hourly wage analysis.

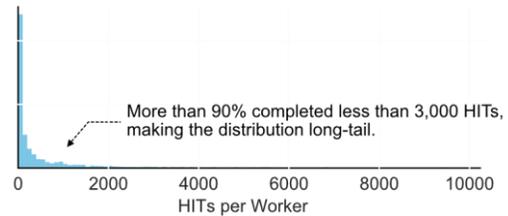

Figure 2. Histogram of performed HIT counts by workers.

On average, workers worked on 1,302 HITs each (SD=4722.5; median=128.5), spending 54.0 hours on average (SD=172.4; median=6.14h). Figure 2 shows the distribution of the total number of HITs completed by workers. One worker completed 107,432 HITs, whereas 135 workers completed only one HIT. Workers used the Crowd Worker plugin for 69.6 days on average (SD=106.3; median=25 days).

Some HITs were submitted with abnormally short or long work duration. This could be because these HITs were completed by automated scripts, submitted prematurely or workers abandoned/forgot to perform the tasks. To mitigate the effect of these outliers on our results, we filtered out top and bottom 5-percentile of the submitted HIT records based on their task duration, leaving N=3,471,580 HITs (91.2% of the original number). The remaining data represents 99,056 unique HIT groups, N=2,666 unique workers, and 19,598 unique requesters.

We retain the N=23,268 (0.7%) HITs with $0 reward, which are typically qualification HITs (*e.g.,* answering profile surveys). We keep these tasks in our dataset as time completing these tasks is still work even if it is not rewarded as such by the requesters. The small portion of the records does not significantly impact our results.

**THE AMT WAGE DISTRIBUTION**

In this section, we analyze the level and distribution of hourly wages and earnings on AMT. We first outline a set of methods to calculate hourly wages before reporting detailed descriptive statistics including total and hourly earnings.

**Measuring the Hourly Wage**

Work on AMT is organized and remunerated as a piece rate system in which workers are paid for successfully completed tasks. Our work log record includes $Time_{submit}$, $Time_{accept}$ and the *Reward* for each HIT. If *HIT Interval*=$Time_{submit}$ - $Time_{accept}$ accurately reflected time spent working on a *HIT*, then it would be simple to calculate the hourly wage associated with each task as *Reward / HIT Interval*. (Note that when the worker returns the HIT, we use *Reward*=$0 regardless of the HIT reward.) Similarly, we could calculate the average per-worker hourly wage with $\sum Reward / \sum HIT\ Interval$ – sum of the total

reward over the total HIT duration that a person earned/spent over the course of working on HITs. We refer to this as the *interval-based* method of computing per-HIT and per-worker hourly wage.

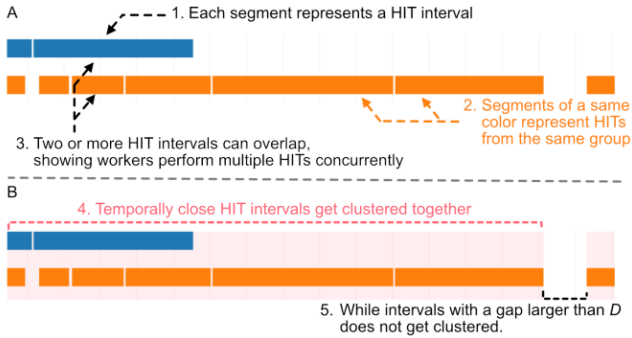

**Figure 3. Timeline visualization of HIT intervals and depiction of the temporal clustering method. The HIT interval data comes from one of the workers in our dataset.**

But the *HIT Interval* does not always correspond directly to work time. As depicted in Figure 3a, HIT Intervals can overlap when a worker accepts multiple HITs at once, and then completes them one-by-one. This is a common strategy that workers use to secure the HITs that they want to work on to prevent them from being taken by other workers. This could cause the interval-based method to underestimate the hourly wage because any time lag between accepting a HIT and starting to work on it will be counted as work time.

There is also a question over how to treat the time between HITs when calculating the hourly wage. When a worker works on HITs in the same HIT group or looks for a new HIT using AMT's search interface, there can be a lag between submitting one HIT and accepting the next. This seems important to count as part of working time but is not captured by the interval-based method, which could lead the interval-based method to overestimate the hourly wage.

To take into account overlapping HITs and the time between tasks, we needed to temporally cluster the HITs into contiguous working sessions. We used a temporal clustering method following Monroe *et al.* [52] that groups a series of temporally close time intervals into clusters using an interval threshold, $D$. For example, given a pair of HITs that are sorted by $Time_{accepted}$, the algorithm will group these HITs into a single cluster if the duration between the first HIT's $Time_{submitted}$ timestamp and the second HIT's $Time_{accepted}$ is smaller than $D$—see Figure 3b. Then, the cluster's last $Time_{submitted}$ is compared with the subsequent HIT. If the duration between the next HIT's $Time_{accepted}$ timestamp is smaller than gap $D$, the algorithm puts the HIT into this cluster. Otherwise, the subsequent HIT forms a new cluster. We call this the *cluster-based* method of measuring the hourly wage.

Different choices of $D$ yield different estimates of working time and thus hourly wages. With $D$=0, only concurrently occurring HITs are clustered together. We also report results for a choice of $D$>0. With $D$>0, HITs that are worked on sequentially but with slight intervals between submitting one task and accepting the next are clustered. Figure 4 shows how the number of clusters in the data set varies with $D$. The Elbow point is 1min [41]—the change in the number of clusters formed diminishes sharply after $D$=1min. This seems sensible as most intervals between submitting and accepting HITs within the same work session should be small. Thus, in addition to the interval-based method, we report wage results using the cluster-based method with $D$=0min and $D$=1min. We compute the *per-cluster hourly wage* for a cluster $C$ as:

$$w_C = \sum_{t \in C} Reward_t / (max_{t \in C}\{Time_{submit,t}\} - min_{t \in C}\{Time_{accept,t}\}) \quad (Eq.\ 1)$$

where $t$ refers to a task. The per-worker average hourly wage is then calculated as $w = \sum \delta_C w_C$ where $\delta_C$ is the fraction of time spent on cluster $C$ relative to all time spent working.

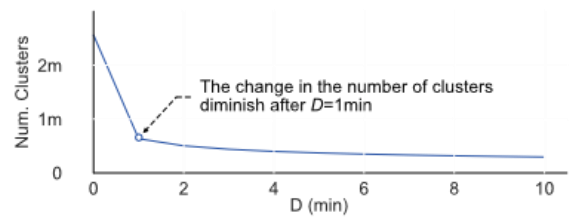

**Figure 4. Line chart of the number of clusters formed. The change in the number becomes small after $D$=1min.**

### Hourly Wages per HIT/Cluster

We first report statistics on effective wage rates at the task level, calculated using our three different methods. In summary, depending on the measure used, mean wage rates per work-unit vary between $4.80/h and $6.19/h. N=600,763 (23.5%) of 0min clusters generated an hourly wage of $7.25, whereas N=80,427 (12.7%) of 1min clusters generated above the federal minimum wage. Table 1 gives the relevant summary statistics.

|  | Per-HIT/Cluster ($/h) | | |
| --- | --- | --- | --- |
|  | Median | Mean | SD |
| **Interval** (N=3,471,580) | 2.54 | 5.66 | 24.1 |
| **Cluster (D=0; N=2,560,066)** | 3.18 | 6.19 | 26.4 |
| **Cluster (D=1; N=635,198)** | 1.77 | 4.80 | 43.4 |

Figure 5 shows the distribution of per-HIT/cluster hourly wages using the different methods for hourly wage computation, disregarding worker identity. The distributions are zero inflated, because N= 460,939 paid $0, either because they were qualification tasks and/or returned. After removing the $0 HITs, the median hourly wage using the interval-based method is $3.31/h and the mean hourly wage is $6.53/h (SD=25.8). We will revisit the impact of the returned HITs to the worker income later.

At *D*=0, N=2,560,066 clusters were formed. N=2,429,384 had only 1 HIT in a cluster—*i.e.,* 70% of HITs were not overlapping. Overlapping HITs came from N=1,629 workers. This indicates that 38.9% of the workers never worked on HITs in parallel and 61.1% of the workers work on two or more HITs in parallel. Taking into account the overlapping nature of tasks raises estimates of average work-unit wage rates as shown in Figure 5a&b.

At *D*=1, N=635,198 clusters were formed. The median and mean per-cluster hourly wages were $1.77/h and $4.80/h (SD=43.4) (Figure 5c). N=331,770 had only 1 HIT in a cluster. Compared to the statistics in case of *D*=0, the mean and median per-cluster hourly wages dropped by 1.39 and 1.41. This indicates that the unpaid time intervals between accepting and submitting HITs have a non-negligible amount of impact to the hourly wage of the workers.

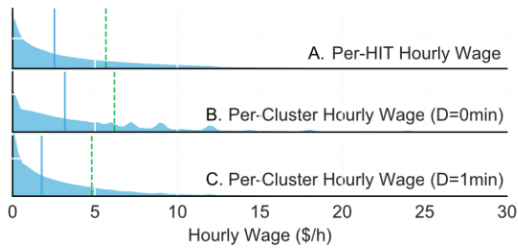

**Figure 5. Distributions of per-HIT and per-cluster hourly wages. The blue and green lines indicate median and mean.**

**Hourly Wages per Worker**
Average hourly wages per worker are lower than those at the task/cluster level. This is because small number of workers are contributing a large number of high hourly wage HITs. Depending on the method used, mean hourly wages per worker on AMT lie between $3.13/h and $3.48/h, while the median wage lies between $1.77/h and $2.11/h—see Table 2. Only 4.2% of workers earn more than the federal minimum wage on average.

| | Per-Worker ($/h) | | |
|---|---|---|---|
| | **Median** | **Mean** | **SD** |
| **Interval** | 1.77 | 3.13 | 25.5 |
| **Cluster (D=0)** | 2.11 | 3.48 | 25.1 |
| **Cluster (D=1)** | 1.99 | 3.32 | 25.0 |

**Table 2. Summary of per-worker hourly wage statistics.**

Figure 6 shows the distribution of the per-worker hourly wage and Table 2 gives the relevant summary statistics. On average, the workers earned $95.96 (SD=310.56; median=$11.90). Compared to the interval-based per-worker hourly wage, cluster based median wages are 19.2% (=2.11/1.77) and 12.4% (1.99/1.77) larger for D=0min and D=1min respectively. This indicates that the workers are benefiting from working in parallel, to some extent.

The wage distributions are positively skewed, with a small proportion earning average wages in excess of $5/h. There are N=111 (4.2%) workers who are making more than minimum wage according to the interval-based method. The number of HITs performed by these workers ranged from 1 to 94,608 (median=12, mean=1512.8, SD=9586.8). Thus, we cannot attribute the high-hourly wage to experience on the platform alone, which does not explain the high hourly wage of more than half of the workers who completed N=12 tasks or less. To further investigate *why* these workers are earning more, we investigate the factors affecting low/high hourly wage in the next section. We use the interval-based method to compute hourly wage unless otherwise noted, because (i) the clustering methods for calculating wages does not provide granular task-level hourly wage information that is necessary in some of the analyses below and (ii) the interval-based method does not over/underestimate the wage much.

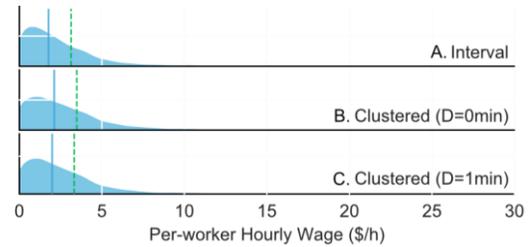

**Figure 6. Distributions of per-worker hourly wages based on the interval-based and cluster-based methods. The blue and green lines indicate median and mean.**

**FACTORS AFFECTING THE HOURLY WAGE**
In this section, we analyze the effect of (i) unpaid work, (ii) HIT reward, (iii) requester behaviors, (iv) qualifications, and (v) HIT type on the hourly wage to identify potential strategies for workers to increase their earnings.

**Unpaid Work**
It is not always obvious what counts as work on crowdsourcing platforms. Working on AMT often involves *invisible* work [62]—time spent on work that is directly/indirectly related to completing HITs yet unpaid. There are several types of this invisible work, including the time spent on the returned HITs, work done for the rejected HITs, and, again, time spent searching for HITs [26,49,51]. While these issues have been identified in prior work, their significance to hourly wage is not quantified. Below, we look into the impact of *returned HITs, rejected HITs,* and *time between HITs* on worker hourly wages.

*Returned HITs*
Of the 3.5m HITs, N=3,027,952 (87.2%) were submitted and N=443,628 (12.8%) were returned. For the submitted HITs, the median and mean work durations were 41s and 116.8s (SD=176.4s). For the returned HITs, the median and mean time spent were 28.4s and 371.5s (SD=2909.8). The total work duration of submitted and returned HITs was 143,981 hours. 98,202 hours were spent on the submitted HITs and 45,778 hours were spent on the returned HITs.

We cannot quantify exactly how much monetary value has been lost due to the 12.8% of the work that was never compensated. However, if we assume that the workers could have earned $1.77/h or $7.25/h—the interval-based

hourly wage and the U.S. minimum wage—$81,027 (1.77 x 45,778) or $331,890 (7.25 x 45,778) was unpaid.

On average, each worker in our dataset returned 26.5% of HITs and spent 17.2 hours on average (SD=71.7, Median=0.9 hours) on them. Evaluating these tasks at the hourly wage ($1.77/h) suggests that workers wasted $30.44 worth of time on average. This shows that returning HITs introduce a significant amount of monetary loss.

In our dataset we cannot observe *why* a worker returns a HIT. So investigating why HITs are returned should thus be a key area of future research; it could be because of poor task instructions that prohibits workers from completing a HIT, broken interface that prohibits submitting HITs, a worker not enjoying a task, and others.

*Rejected HITs*
In our dataset, N=1,029,162 out of 3.5m HITs (29.6%) had *'approved'* or *'rejected'* status. Within these records, N=1,022,856 records (99.4%) were approved and N=6,306 (0.6%) were rejected. In terms of total time spent on the approved and rejected HITs, 33,130 hours were spent on the approved HITs (99.3%) and 240 hours were spent on the rejected HITs (0.7%).

This suggests that, at least within the scope of our data, HIT rejection is a smaller issue in terms of unpaid work as nearly 100% of work was accepted. Note, however, as McInnis revealed [51], workers are sensitive to rejection because a poor approval rate could prohibit them from doing some tasks on the market (because some HITs require a high approval rate) or get them banned permanently from the platform. Avoiding rejection (*e.g.,* by returning) could be contributing to the high acceptance rate.

*Time between HITs*
Our cluster-based analysis of the worker hourly wage suggests that there is non-negligible amount of unpaid time spent between HITs. Some portion of this likely represents time taken for searching for HITs, waiting for a page to load or accepting new HITs, although we have no way to know how workers are spending their time between HITs. We now investigate the effect of the unpaid time between HITs on the worker hourly wage. We do this by computing the total cluster-based task durations with *D*=1min and *D*=0min, and subtracting the former by the latter.

In total, workers spent 103,030 hours working according to the *D*=1min cluster-based total duration and 98,427 hours working based on the *D*=0min cluster based duration. This implies that 4602.7 hours were spent between HITs. Time spent between HITs sums up to 103.6 minutes on per-worker average. Median total time between HITs per worker was 12.1 minutes. Naturally, people who worked on more HITs had larger unpaid time between HITs. Using the median hourly wage ($1.77/h) and the US federal minimum wage ($7.25), the monetary values of the total unpaid time amounts to $8,146.78 and $33,369.58.

*Takeaway*
Returning HITs has the biggest impact to the hourly wage. The time lost due to the time between the HITs has the second most impact. Task rejection has the least impact in terms of unpaid work. Note, however, rejection could have potential risks of not being able to accept HITs in the future or getting banned from AMT, which is not quantified here.

**HIT Reward**
The hourly wage depends both on HIT reward *and* how long it takes to complete a task. While it might seem obvious that higher reward HITs result in higher hourly wages, this relationship might not hold if higher paying HITs take proportionately longer to complete.

To investigate the relationship between HIT reward and hourly wages, we examine the association between the mean interval-based HIT hourly wage of HIT groups and HIT reward. Similar to the per-worker hourly wage, mean per-group hourly wage is $\sum Reward / \sum HIT\ Interval$, summed over the tasks in the same HIT group. In the analysis of this section, we omit the HITs that had $0 reward to remove the effect of returned HITs.

As the HIT reward distribution is highly skewed, we apply the Box-Cox transformation[1] ($\lambda$=0.174) for our analysis. We select this transformation over alternatives as it generated better fit for the regression model that we describe in the next paragraph. We transform the per-HIT group hourly wage using the log transformation.

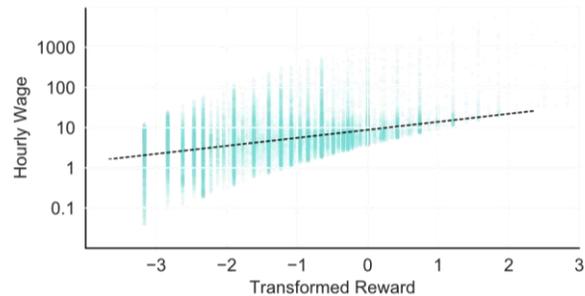

**Figure 7. The scatter plot showing the relationship between the transformed reward and hourly wage. The line represents the model that we fit with ordinary linear regression.**

We fit the transformed reward and hourly wage using ordinary least squares regression. The residuals are slightly skewed and peaked compared to the normal distribution assumed by the model (*skew*=0.7, *kurtosis*=2.2), but it should be tolerable. We obtain $y = 1.996x + 0.9465$ ($R^2$=0.18; Figure 7). The obtained linear model is for the transformed data. In the original dimension, this model suggests that if a worker completes a HIT with $0.01 reward, they should expect to earn $2.06/h. Similarly, working on a HIT with $1.00 reward should yield $8.84/h

---

[1] Box-Cox transform is given by $y = \frac{x^\lambda - 1}{\lambda}$ [20]

and working on HITs with rewards above $0.64 should pay workers above the minimum wage ($7.25/h).

We point out that there are low-reward HITs that yield high-hourly wage, too. This means some low reward HITs are indeed priced fairly in terms of the time that a worker has to spend to complete them. It is, however, harder for workers to distinguish low-reward HITs that yield high hourly wage and low hourly wage *a priori*.

*Takeaway*
High reward HITs yield a higher hourly wage, indicating that while they take longer to perform, they do *not take so much longer* as to eliminate the gains of the higher piece rate. The analysis suggests an easy-to-employ strategy for worker to increase their wages (*i.e.,* take high reward HITs).

**Requesters**
Workers seek to find good requesters so they can earn fairer rewards [49,51]. To do so they use tools such as Turkopticon [38] and information from sites like Turker Nation [67]. In this section, we evaluate how much variation in hourly wages there is across requesters.

We use the interval-based method to compute the per HIT hourly wage—or *hourly payment* from the requester's perspective—and in turn per-requester hourly payment. The other methods for calculating wages do not make sense for this analysis because tasks grouped together may come from different requesters. Overall, there were N=19,598 requesters who posted at least 1 HIT in our dataset. On average, requesters posted N=173.4 HITs (SD=4283.0), with the median N=6 HITs. To investigate the characteristics of actual payments, we removed the HITs that were returned. This reduces the HITs to 3.03 million records from N=16,721 requesters (*i.e.,* HITs in our records from 2,839 requesters were never completed). We also filter out qualification HITs that had $0 reward. Per-HIT level hourly payment follows the same trend as what we saw in the analyses of interval-based hourly *wage*, so we skip that analyses and only report the per-requester statistics.

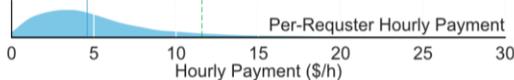

**Figure 8. KDE plot of per-requester hourly payment.**

Using the filtered data, we computed the hourly payment per requester. On average, requesters paid $11.58/h (SD=145.71; median=$4.57/h)—see Figure 8. Mean and median are higher than per-worker hourly wage statistics (*e.g.,* interval-based mean and median wages are $3.13/h and $1.77/h). This suggests that the large sum of low-paid HITs is posted by a relatively small number of requesters.

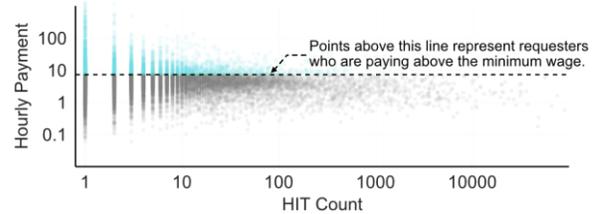

**Figure 9. A scatter plot of per-requester HIT count *vs.* per-requester hourly payment.**

The aggregate statistics disregard the number of HITs the requester posted. For example, 3,667 requesters posted only 1 HIT, whereas 1 requester posted 405,709 HITs. Therefore, we created a scatter plot that depicts the relationship between per-requester hourly payment and how many HITs the requesters posted (Figure 9).

In Figure 9, each dot represents a requester. The x-axis is the number of HITs posted by each requester, while the y-axis represents per-requester hourly payment. The dashed line indicates 7.25/h. The green points above the dashed line indicate the requesters who paid more than minimum wage (N=4,473). N=962 of them posted more than 10 HITs and N=3,511 posted less than 10 HITs. While many requesters post a large amount of low-payment HITs, there are requesters who are posting above median number of HITs that yield fair wage. This validates that it is feasible to get fair hourly wage if you can find good requesters.

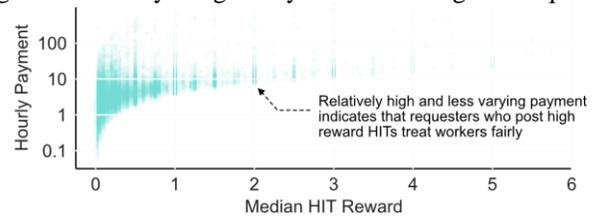

**Figure 10. A scatter plot of median HIT reward paid by requesters *vs.* per-requester hourly payment.**

Do requesters who post high-reward HITs pay more fairly? To validate this, we look into the relationship between the per-requester HIT reward and hourly payment. Figure 10 shows the median HIT reward per-requester on the x-axis and the hourly payment on the y-axis. The graph indicates that working for requesters who constantly post high reward HITs is associated with earning a higher hourly wage. This corresponds to the insight from the previous analysis that working on high reward HITs can be lucrative.

*Takeaway*
Though the majority of the requesters pay below minimum wage, we saw that there are requesters who are fair. They post a significant number of HITs. Thus, finding these requesters and giving their work priority could improve worker hourly wage. Existing worker tools could support this by watching for the presence of HITs from these requesters and alerting workers to their presence [68].

## Qualifications

Qualifications allow requesters to selectively hire workers. As this gives requesters the potential to hire only skilled and/or targeted workers, it is plausible that HITs that require qualification pay more generously. We thus compare the wage of HITs with and without qualification requirements.

Our dataset contains N=1,801 unique qualification types. N=36,068 unique HIT groups requires at least one qualification, which corresponds to N=1,711,473 HITs. N=1,760,107 HITs did not require qualifications. The median interval-based hourly wage of the HITs with and without qualification requirements were $2.63/h and $2.45/h respectively. Likewise, means were $5.65/h (SD=19.1) and $5.67/h (SD=28.2). Unlike what we expected, the wages did not differ much between groups.

| Qualification Name | HIT Count | Median ($/h) | Mean ($/h) | SD |
|---|---|---|---|---|
| HIT approval rate (%) | 1309320 | 2.30 | 5.55 | 17.16 |
| Location | 937701 | 3.10 | 6.13 | 19.89 |
| Total approved HITs | 647831 | 4.14 | 7.80 | 19.27 |
| Adult Content Qualification | 410193 | 4.25 | 6.05 | 7.76 |
| Category Validation | 207581 | 3.81 | 4.99 | 4.57 |
| Blocked | 145782 | 4.19 | 5.44 | 4.90 |
| HIT abandonment rate (%) | 83145 | 2.47 | 4.89 | 7.09 |

Table 3 summarizes the ten most common qualification types. For example, N=1,309,320 HITs required the workers to have "HIT approval rate (%)" qualification, N=937,701 HITs required the workers to have "Location" qualification, and so on. Some qualifications seem to correspond to higher hourly wage. Figure 11 shows that 7 out the ten most common qualifications such as "Total approved HITs" and "Question Editor" corresponded to higher hourly wage compared to the overall average wage.

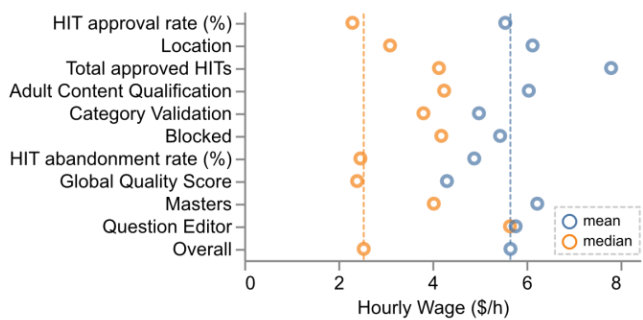

**Figure 11. Mean and median hourly wages yielded by the HITs that require the ten most common qualification types.**

*Takeaway*
The HITs that require qualifications do not necessarily yield a higher hourly wage. But there are some qualifications that correspond to higher hourly wage (*e.g.,* "Question Editor").

## Types of Work

What *types or topics* of HITs yield high hourly wage? Knowing that could guide workers to selectively work on types of work associated with high-wage. Unfortunately, the platform does not provide adequate information about a topic of a particular HIT. While requesters could provide HIT keywords, they are not necessarily consistent across HITs and keywords could be ambiguous. In fact, in our pilot analysis where we studied the relationship between keyword and hourly wage, keyword ambiguity prevented us from understanding what topic/keyword of HITs yield high hourly wage (*e.g.,* audio and image 'transcription' gets mixed up). Thus keywords alone were not adequate to characterize HIT types, which led us to also take information from HITs' titles and descriptions.

Manually labeling HITs' topics based on HIT title, description, and keywords—we call this triplet a *HIT document*—cannot be done by any one person because of the data volume. We thus turn to a labeling process mediated by computation known as *topic modeling*. Topic modeling methods are often used to categorize a set of unlabeled documents into a finite number of coherent topic clusters [7,28,63], in turn helping researchers to navigate a large corpus (HIT documents in our case) [17,18,24,54]. However, unsupervised clustering of unlabeled documents is far from a solved problem [16,18]. Although there are existing machine learning algorithms like K-Means [32] and Latent Dirichlet Allocation [7], fully automated algorithm still cannot cluster documents with adequate accuracy.

The lack of a go-to method for document clustering necessitates us to iteratively explore different methods that allow us to efficiently and effectively categorize HIT documents into a set of topics. After trial-and-errors, we settled in using a three-step method, which involved: (i) transform each HIT document into vector representation using *term frequency-inverse document frequency* (*TF-IDF*) and *Latent Semantic Analysis* (*LSA*) transformations; (ii) retrieve a list of topical keywords by a semi-automated process that involves automated clustering of the transformed HIT documents and manual process of retrieving recurring topical phrases; and (iii) we use the retrieved topical keywords to *query* HITs in our dataset.

*Vector Representation of HIT Description*
The starting point of the process is preprocessing of HIT descriptions. From each HIT group, we extract requester-specified title, description, and keywords, then concatenate them into a single *document*. To suppress the effect of uninformative words to the subsequent steps, we remove stop words (*e.g.,* "the"), remove special characters (*e.g.,* "*"), turn them into lower-case, and remove the last character of a term if it ends with "s".

We then create a table of term counts in which rows represent documents, columns represent terms in of the documents, and each cell contains the count of words used in a HIT document. At this point, there is no distinction between the words' *importance* (*e.g.,* terms "survey" and "good" have same weight even though the former is likely more informative for characterizing a HIT topic). A common step for assigning importance to words is TF-IDF transformation [56]. The method assigns a weight to a term that occurs frequently in a document, but negatively weights the term that appears across documents.

A tabulation with term counting and TF-IDF transformation yields a sparse, high-dimensional matrix. This causes the poorly generalizable representation and negatively impacts the subsequent clustering and querying steps. We thus use Latent Semantic Analysis which maps a sparse vector representation of a document into denser latent vectors [58]. Following the advice by Evangelopoulos *et al* [23], we map the rows of HIT documents into $k_{dim}$=300 dimensions.

*Retrieving HIT Topics*
Given the latent representation of the documents (*i.e.,* HIT group descriptions), we move on to retrieving the topical keywords. This sub-process includes two steps: (i) K-Means based automated document clustering and (ii) manual retrieval of topical phrases from the clustered documents.

We use K-Means algorithm to cluster the transformed documents that are close to each other in the latent space [53]. K-Means clustering requires a user to specify a distance function and a number of clusters *a priori*. We use cosine distance to measure the similarity between two documents. Cosine distance returns a value between [0, 2], where 0 indicates that two documents are close or *similar*. We follow Bradford recommendation and use a clusters size
$k_{cluster}$=200 [9].

From each of the formed HIT document 200 clusters, we sampled 50 HIT groups uniformly randomly. We manually go through them and retrieve recurring *topical keywords* (*e.g.,* {"transcribe", "image"}). While 200x50=10k is a large number, this way of retrieving topical phrases is easier compared to randomly going through all 99k HIT groups because clustering algorithm returns some sets of HIT documents with clean, coherent topics. The retrieved topical keywords are listed on Table 4, which we further group into classes of HIT taxonomy given by Gadiraju *et al.* [25] (with an additional category *Research*).

*Querying HITs*
Given the document-to-vector mapping and the list of topical keywords, we can move on to *querying* HITs from the dataset. Using the same transformation that is used to map a HIT document to a latent vector, we map every topical keyword (*e.g.,* {"psychology", "survey"}, {"audio", "transcription"}) into a vector of real values. This allows us to use cosine distance to measure similarity between a topical phrase, which acts as a *search query* and documents. While there is no universally accepted distance threshold for the cosine distance, $d_{cosine}$=0.4 is considered as a good choice [23], which we follow. Although Evangelopoulos *et al.* warns that this threshold is solely based on heuristics, manual inspection of query results validated that relevant documents are returned.

*Query Result*
Figure 12a shows the distributions of hourly wage for the seven HIT categories. Duplicate HITs that are associated with two or more topical phrases are dropped for aggregate statistic computation and plotting the distribution. For each of Information Finding (IF), Verification and Validation (VV), Interpretation and Analysis (IA), Content Creation (CC), Surveys, Content Access (CA), and Research, tuples of (*class*, *mean wage, median wage*) are: (IF, 8.43, 3.78), (VV, 6.78, 3.60), (IA, 11.36, 8.94), (CC, 2.13, 1.26), (Surveys, 9.30, 4.88), (CA, 7.59 , 5.99), and (Research,

| **Information Finding (IF; N=26,203)** |
|---|
| data collection; data extraction; find company name; find contact; find email; find phone number; find url; find website |
| **Verification and Validation (VV; N=13,081)** |
| audio quality assurance; detect spam; verify image; website test |
| **Interpretation and Analysis (IA; N=72,932)** |
| brand evaluation; evaluate brand; image categorization; rate article; rating image; rating picture; rating product; review article; review video; video evaluation; web page categorization |
| **Content Creation (CC; N=320,220)** |
| audio transcription; describe image; describe video; logo design; photo tagging; transcribe audio; transcribe data; transcribe data; transcribe image; transcribe text; translation |
| **Surveys (N=47,192; the term 'survey' is omitted from each phrase for brevity)** |
| academic; activitie; advertising; attitude; behavior; behavioral; belief; brand; college student; consumer behavior; consumer choice; consumer experience; consumer preference; consumer; consumer topic; current event; decision making; decision; demographic; everyday life; game; habit; health; life event; life experience; life; marketing; mental health; opinion; personal; personality; policy; politic; political attitude; preference; product evaluation; product; psychological; psychology; public opinion; public policy; relationship; research; scenario; search result; shopping; smartphone app; social attitude; social experience; social media; social psychology; technology; workplace |
| **Content Access (CA; N=995)** |
| content viewing; image viewing |
| **Research (N=433)** |

7.63, 5.40). CC's hourly wage distribution is highly skewed toward low-wage. IF, VV, and Survey HITs' wages are skewed toward low-wage too, but less so compared to the CC's distribution. Interpretation and Analysis (IA), Content Access (CA), and Research are more flatly distributed, showing HITs of these topics tend to yield higher wage.

Figure 12b shows strip plots where each circle represents a topical query. The size of circles corresponds to the number of HITs retrieved. The x-axis represents the median HIT hourly wage among the retrieved HITs. We observe two HIT groups with large quantities of tasks under CC are pulling down the hourly wage distribution. They correspond to topical keywords {"transcribe", "data"} (N=313,559; median=$1.24/h) and {"transcribe", "image"} (N=152,031;

median=1.13/h). On the other hand, topical phrase {"video", "evaluation"} under the category IA is associated with higher median hourly wage ($10.30/h) and has large quantity (N=49,720), making the group's distribution flat.

*Takeaway*
There is variation in hourly wage between different topics of HITs. We showed that HITs such as data/image transcription are low-paying whereas "video evaluation" HITs are high-paying.

**DISCUSSION**

such tools. Although the Crowd Worker plugin does not collect more information than it requires to compute hourly wages, this may not be visible to workers which could limit adoption of the technology [57].

Unpaid work is an important factor driving low hourly wages on AMT. Workers are not paid when searching for tasks and are not paid for qualification tasks nor tasks that are returned or rejected. We suspect that experienced workers learn over time to minimize such unpaid work. However, encoding this process into a system that can be

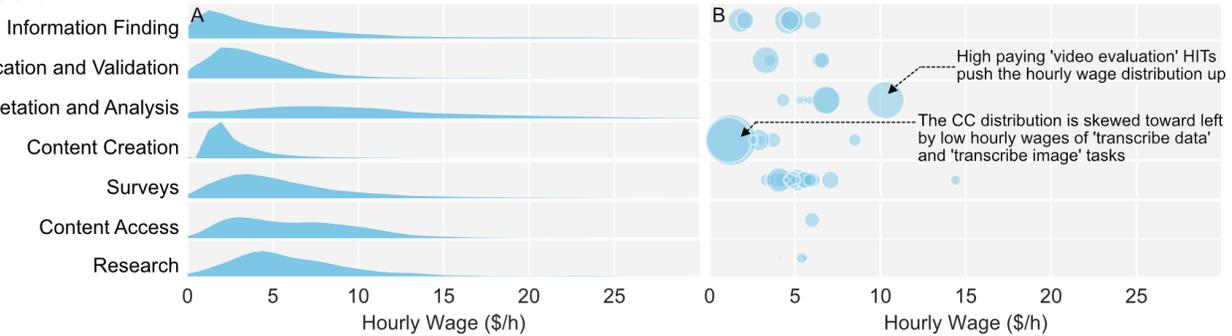

**Figure 12.** (a) Hourly wage distributions of seven HIT categories provided by Gadiraju *et al.* [25] (with an additional category *Research*). (b) Strip plots showing median hourly wages of HITs associated with the topical keywords in Table 4.

We estimate that 96% of workers on AMT earn below the U.S federal minimum wage. While requesters are paying $11.58/h on average, dominant requesters who post many low-wage HITs like content creation tasks are pulling down the overall wage distribution. This is problematic as the primary goal of workers is income generation [4,13,46,49], rather than having fun or making pocket money [2,4,43]. Many people working on low paying HITs are likely from groups traditionally excluded from the formal labor market [1,4,10,64], such as people with disabilities who have challenges in securing jobs at contemporary office work environment [4,65]. Hara and Bigham noted that some crowd work like image transcription can be done by autistic people—a population that has challenge in securing jobs compared to those without disabilities [10,30,33]—but this type of work is exactly what generates the lowest hourly wages as we showed in the topical analysis. We here discuss the implications of our results for the design of worker tools and platform infrastructure, and call on the HCI community to advance research in tools that can help to achieve a fairer distribution of earnings from crowd work.

Raising workers' awareness of their effective hourly wage and suggesting real-time strategies to increase their earnings may help workers to optimize their task selection and working patterns. Providing visualized, easy to interpret information about earnings provides useful feedback for workers that is not provided by AMT. Although measuring hourly wage is not straightforward [3], we showed that different wage computation methods do *not* result in largely different hourly wage estimates. Privacy concerns must be taken into consideration when designing used by novice workers maximize their wage would be beneficial. Tools that automatically push tasks to workers (*e.g.,* [29,30]) and inform them of how likely the task is to be completable and accepted, combined with real-time information about tasks that exist on the market could thus be useful.

The majority of requesters pay below minimum wage. Helping workers to avoid unfair requesters is one way of dealing with this problem. But fair requesters do not post HITs all the time and a solution to the root problem—the presence of unfair requesters—is needed. We may be able to mitigate the problem by facilitating the communication between workers and/or increasing minimum reward.

Workers cite poor communication with requesters as a major flaw in crowd work platform design [4]. Improving communication channels might make it easier for requesters to identify and fix broken HITs (reducing time spent on returned tasks) and enable crowd workers to bargain collectively. While nudging workers to individually or collectively communicate and negotiate is difficult [42,59], overcoming these barriers can be beneficial for workers.

**LIMITATIONS**
Our results may be biased (*e.g., super turkers* [8] who are already experienced may not be interested in using the plugin). That said, our sample includes 2,676 workers, each of whom completed 1.3k tasks on average. Our estimate of work intervals may not correspond to the time spent actively working on HITs, but this is same for any time-based payment scheme. Our data did not capture what scripts workers may have been using to assist them. Those who use scripts may have been automating some parts of their work and earning more than those who do not. While our analyses suggested methods for increasing wage, we do not argue for causal relationship. Like many log analysis, we lack worker demographics. To better understand *who* are the workers in our dataset, we will conduct an online survey study. Investigation of other ethical issues like overtime work is future work.

## CONCLUSION

We used the log data of 2,676 workers performing 3.8 million tasks on Amazon Mechanical Turk to understand worker hourly wages. Our task-level analysis revealed a median hourly wage of ~$2/h, validating past self-report estimates. Only 4% of workers earn more than $7.25/h, justifying concerns about non-payment of the minimum wage. We characterize three sources of unpaid work that impact the hourly wage (*i.e.,* task search, task rejection, task return). We further explore the characteristics of tasks and working patterns that yield higher hourly wages.

## ACKNOWLEDGEMENTS

Adams gratefully acknowledges funding from the Economic and Social Research Council, Grant ES/N017099/1